\documentclass[aps,12pt,showpacs,preprint,groupedaddress,floatfix]{revtex4}
\usepackage{graphicx}
\usepackage{dcolumn}
\usepackage{bm}
\usepackage{amssymb}
\usepackage{amsmath}

\newcommand{\bff}[1]{{\mbox{\boldmath $#1$}}}

\newcommand{\bra}[1]{\left\langle #1 \right|}
\newcommand{\ket}[1]{\left| #1 \right\rangle}

\begin{document}

\title{Energy Density Functional analysis of shape evolution in N=28 isotones}
\author{Z. P. Li$^{1,2}$}
\author{J. M. Yao$^{1}$}
\author{D. Vretenar$^{2}$}
\author{T. Nik\v si\'c$^{2}$}
\author{H. Chen$^{1}$}
\author{J. Meng$^{3,4,5}$}

\affiliation{$^{1}$School of Physical Science and Technology, Southwest University, Chongqing 400715, China}
\affiliation{$^{2}$Physics Department, Faculty of Science, University of Zagreb, 10000 Zagreb, Croatia}
\affiliation{$^{3}$School of Physics and Nuclear Energy Engineering, Beihang University, Beijing 100191, China}
\affiliation{$^{4}$State Key Laboratory of Nuclear Physics and Technology, School of Physics,
Peking University, Beijing 100871, China}
\affiliation{$^{5}$Department of Physics, University of Stellenbosch, Stellenbosch, South Africa}

\bigskip
\date{\today}

\begin{abstract}
The structure of low-energy collective states in proton-deficient $N=28$ isotones is analyzed
using structure models based on the relativistic energy density functional DD-PC1. The
relativistic Hartree-Bogoliubov model for triaxial nuclei is used to calculate binding energy
maps in the $\beta$ - $\gamma$ plane. The evolution of neutron and proton single-particle
levels with quadrupole deformation, and the occurrence of gaps around the Fermi surface,
provide a simple microscopic
interpretation of the onset of deformation and shape coexistence.
Starting from self-consistent constrained energy surfaces calculated with the functional DD-PC1,
a collective Hamiltonian for quadrupole vibrations and rotations
is employed in the analysis of excitation spectra and transition rates of $^{46}$Ar, $^{44}$S,
and $^{42}$Si. The results are compared to available data, and previous
studies based either on the mean-field approach or large-scale shell-model
calculations. The present study is particularly focused on $^{44}$S, for which data
have recently been reported that indicate pronounced shape coexistence.

\end{abstract}


\maketitle

\section{\label{secI}Introduction}

Shapes of neutron-rich nuclei far from stability
have extensively been explored in many experimental and theoretical studies.
The evolution of ground-state shapes in an isotopic or isotonic
chain, for instance, is governed by changes of the
underlying shell structure of single-nucleon orbitals.  In particular far from the $\beta$-stability line, the
energy spacings between single-nucleon levels change considerably with the number of neutrons or protons.
This can lead to reduced spherical shell gaps, and in some cases spherical magic numbers may
partly or entirely disappear \cite{SorPor08}. The reduction of spherical shell closure often
leads to the occurrence of ground-states deformation and, in a number of cases, to the
coexistence of different shapes in a single nucleus.

In recent years a number of studies have been devoted to the investigation of the fragility of
the $N=28$ magic number in neutron-rich nuclei \cite{OSor.10}. In $\beta$-stable nuclei the
$Z$ or $N =28$ shell closure is the first magic number produced by the spin-orbit part of
the single-nucleon potential, which lowers the $f_{7/2}$ orbital with respect to the $p_{3/2}$
and thus forms a spherical shell gap at nucleon number 28. However, as a number of
experimental investigations have shown \cite{Sorlin93,Scheit96,Glasmacher97,Sohler02,
Gade05,Grevy05,Gaudefroy06,Campbell06,Bastin07,Gaudefroy09,Force10},
in the proton-deficient $N=28$ isotones below $^{48}$Ca the spherical shell gap is
progressively reduced and the low-energy spectra of $^{46}$Ar, $^{44}$S, and $^{42}$Si
display evidence of ground-state deformation and shape-coexistence.

Both large-scale shell model (SM) calculations \cite{Scheit96,Glasmacher97,Retamosa97,Dean99,Gaudefroy10,Sohler02,
Caurier04,Caurier05,Gaudefroy06, Campbell06,Bastin07,Gaudefroy09,Nowacki09,Force10}
and self-consistent mean-field (SCMF) models \cite{Werner96,Hirata96,Glasmacher97,Lalazissis99,
Carlson00,Peru00,Sohler02,Guzman02,Torres10} have been employed in the theoretical
description of these phenomena. The basic advantages of the SM
approach include the ability to  simultaneously describe all spectroscopic
properties of low-lying states,  the use of effective interactions that can be
related to microscopic inter-nucleon forces, and the description of collective
properties in the laboratory frame. On the other hand, since SM
effective interactions depend on the choice of active
shells and truncation schemes, there is no universal shell-model
interaction that can be used for all nuclei.

A variety of structure phenomena, including regions of exotic
nuclei far from the line of $\beta$-stability and close to the nucleon
drip-lines, have been successfully described with mean-field
models based on the Gogny interaction, the Skyrme energy functional,
and the relativistic meson-exchange effective Lagrangian
\cite{BHR.03,VALR.05,Meng06}. The SCMF approach to
nuclear structure enables a description of
the nuclear many-body problem in terms of a universal energy density
functional (EDF). When extended to also take into account
collective correlations, this framework provides a detailed microscopic
description of structure phenomena associated with shell evolution.
Compared to the SM, the strong points
of the mean-field approach are the use of global functionals, the treatment of
arbitrarily heavy systems, model spaces that include all occupied states
(no distinction between core and valence nucleons, no need for effective
charges) and the intuitive picture of intrinsic shapes.

A quantitative description of shell evolution, and in particular
the treatment of shape coexistence phenomena, necessitates the
inclusion of many-body correlations beyond the mean-field
approximation. The starting point is usually a constrained Hartree-Fock
plus BCS (HFBCS), or Hartree-Fock-Bogoliubov (HFB) calculation of the
binding energy surface with the mass quadrupole components as
constrained quantities. In most studies calculations have
been restricted to axially symmetric, parity conserving configurations.
The erosion of spherical shell-closures in nuclei far from stability
leads to deformed intrinsic states and, in some cases, mean-field
potential energy surfaces with almost degenerate prolate and oblate
minima. In order to describe nuclei with soft potential energy surfaces
and/or small energy differences between coexisting minima, it is necessary
to explicitly consider correlation effects beyond the mean-field level.
The rotational energy correction, i.e. the energy gained by the restoration
of rotational symmetry, is proportional to the quadrupole deformation of
the intrinsic state and can reach several MeV for a well deformed configuration.
Fluctuations of quadrupole deformation also contribute to the
correlation energy. Both types of correlations can be included
simultaneously by mixing angular momentum projected states corresponding
to different quadrupole moments. The most effective approach for
configuration mixing calculations is the generator coordinate method (GCM),
with multipole moments used as coordinates that generate the intrinsic wave
functions.

In recent years several accurate and efficient models, based on microscopic
energy density functionals, have been developed that perform restoration of
symmetries broken by the static nuclear mean field, and take into account
quadrupole fluctuations. However, while GCM configuration mixing of axially
symmetric states has routinely been employed in structure studies, the application
of this method to triaxial shapes presents a much more involved and technically
difficult problem. Only the most recent advances in parallel computing and modeling
have enabled the implementation of microscopic models, based on triaxial
symmetry-breaking intrinsic states that are projected on particle number and angular
momentum, and finally mixed by the generator coordinate method \cite{Bender08,Rodriguez10,Yao10,NVR.11}.

In an approximation to the full GCM for five-dimensional quadrupole dynamics,
a collective Hamiltonian can be formulated that restores rotational symmetry and accounts
for fluctuations around mean-field minima. The dynamics of the five-dimensional
Hamiltonian for quadrupole vibrational and rotational degrees of freedom is governed by the
seven functions of the intrinsic deformations $\beta$ and $\gamma$ : the collective potential,
the three vibrational mass parameters, and three moments of inertia for rotations around the principal
axes. These functions are determined by microscopic mean-field calculations using a
universal nuclear EDF. Starting from self-consistent single-nucleon orbitals, the corresponding occupation
probabilities and energies at each point on the constrained energy surfaces, the mass parameters and the
moments of inertia are calculated as functions of the deformations $\beta$ and $\gamma$.
The diagonalization of the resulting Hamiltonian yields excitation energies and collective wave
functions that can be used to calculate various observables, such as electromagnetic transition
rates \cite{Ni.09,Li09a}.
In this work we employ a recent implementation of the collective Hamiltonian for quadrupole degrees
of freedom in a study of shape coexistence and low-energy collective states in $N=28$ isotones.

Both non-relativistic and relativistic energy density functionals have been used in SCMF studies
of the erosion of the $N=28$ spherical shell gap. One of the advantages of using relativistic
functionals, particularly evident in the example of $N=28$ isotones, is the natural inclusion of the
nucleon spin degree of freedom, and the resulting nuclear spin-orbit potential
which emerges automatically with the empirical strength in a covariant formulation. In the present
analysis we use the new relativistic functional DD-PC1 \cite{NVR.08}. Starting from microscopic
nucleon self-energies in nuclear matter, and empirical global properties of the nuclear matter
equation of state, the coupling parameters of DD-PC1 were fine-tuned to the experimental
masses of a set of 64 deformed nuclei in the mass regions $A \approx 150-180$ and
$A \approx 230-250$. The functional has been further tested in calculations of medium-heavy
and heavy nuclei, including binding energies, charge radii, deformation parameters, neutron skin
thickness, and excitation energies of giant multipole resonances. The present calculation of
$N=28$ isotones, therefore, presents an extrapolation of DD-PC1 to a region of nuclei very
different from the mass regions where the parameters of the functional were adjusted, and
thus a test of the global applicability of DD-PC1.

Section \ref{secII} includes a short review of the theoretical framework: the relativistic
 Hartee-Bogoliubov model for triaxial nuclei, and the corresponding
 collective Hamiltonian for quadrupole degrees of freedom. The evolution of shapes
 in the $N=28$ isotones is analyzed in Sec.~\ref {secIII}: the quadrupole constrained energy
 surfaces determined by DD-PC1, and the resulting low-energy collective spectra, in comparison
 to available data and previous SCMF and SM calculations.
 Section \ref{secIV} summarizes the results and ends with an outlook for future studies.
\section{\label{secII} Theoretical framework}
\subsection{3D relativistic Hartee-Bogoliubov model with a separable
pairing interaction}
The relativistic Hartee-Bogoliubov model \cite{VALR.05,Meng06}
provides a unified description of particle-hole $(ph)$ and particle-particle
$(pp)$ correlations on a mean-field level by combining two average
potentials: the self-consistent mean field that
encloses long range \textit{ph} correlations, and a
pairing field $\hat{\Delta}$ which sums up
\textit{pp}-correlations. In the present analysis the
mean-field potential is determined by the relativistic density functional
DD-PC1 \cite{NVR.08} in the $ph$ channel, and a new separable pairing
interaction, recently introduced in Refs.~\cite{TMR.09a,Niksic10}, is used
in the $pp$ channel.

In the RHB framework the mean-field state is described by a generalized Slater determinant
$|\Phi\rangle$ that represents the vacuum with respect to independent
quasiparticles. The quasiparticle operators are defined by the
unitary Bogoliubov transformation,  and the corresponding  Hartree-Bogoliubov wave
functions $U$ and $V$ are determined by the solution of the RHB equation.
In coordinate representation:
\begin{equation}
\label{eq:RHB}\left(
\begin{array}
[c]{cc}%
h_{D}-m-\lambda & \Delta\\
-\Delta^{*} & -h_{D}^{*}+m+\lambda
\end{array}
\right)  \left(
\begin{array}
[c]{c}%
U_{k}({\mbox{\boldmath $r$}})\\
V_{k}({\mbox{\boldmath $r$}})
\end{array}
\right)  = E_{k} \left(
\begin{array}
[c]{c}%
U_{k}({\mbox{\boldmath $r$}})\\
V_{k}({\mbox{\boldmath $r$}})
\end{array}
\right)  \;.
\end{equation}
In the relativistic case the self-consistent mean-field corresponds to the
single-nucleon Dirac Hamiltonian $\hat{h}_{D}$,
$m$ is the nucleon mass, and the chemical potential
$\lambda$ is determined by the particle number subsidiary condition such
that the expectation value of the particle number operator in the ground state
equals the number of nucleons. The pairing field $\Delta$ reads
\begin{equation}
\Delta_{ab}({\mbox{\boldmath $r$}},{\mbox{\boldmath $r$}}^{\prime}) = \frac
{1}{2}\sum_{c,d}{V_{abcd}({\mbox{\boldmath $r$}},{\mbox{\boldmath $r$}}%
^{\prime})} \kappa_{cd}({\mbox{\boldmath $r$}},{\mbox{\boldmath $r$}}^{\prime
}).
\end{equation}
where $V_{abcd}({\mbox{\boldmath $r$}},{\mbox{\boldmath $r$}}^{\prime})$ are
the matrix elements of the two-body pairing interaction, and the indices $a$,
$b$, $c$ and $d$ denote the quantum numbers that specify the Dirac indices of
the spinor. The column vectors denote the quasiparticle wave functions, and
$E_{k}$ are the quasiparticle energies.

The single-particle density and the pairing tensor, constructed from the
quasiparticle wave functions
\begin{align}
\label{eq:pairing-tensor}\rho_{cd}({\mbox{\boldmath $r$}}%
,{\mbox{\boldmath $r$}}^{\prime})  &  = \sum_{k>0}{V^{*}_{ck}%
({\mbox{\boldmath $r$}})V_{dk}({\mbox{\boldmath $r$}}^{\prime})},\\
\kappa_{cd}({\mbox{\boldmath $r$}},{\mbox{\boldmath $r$}}^{\prime})  &
=\sum_{k>0}{U^{*}_{ck}({\mbox{\boldmath $r$}})V_{dk}({\mbox{\boldmath $r$}}%
^{\prime})},
\end{align}
are calculated in the \emph{no-sea} approximation (denoted by $k>0$): the
summation runs over all quasiparticle states $k$ with positive quasiparticle
energies $E_{k}>0$, but omits states that originate from the Dirac sea. The
latter are characterized by quasiparticle energies larger than the Dirac gap
($\approx1200$ MeV).

In most applications of the RHB model the pairing part of the
Gogny force~\cite{BGG.91} was used in the particle-particle ($pp$) channel.
A basic advantage of the Gogny force is the
finite range, which automatically guarantees a proper cut-off in momentum
space. However, the resulting pairing field is non-local and the solution of
the corresponding Dirac-Hartree-Bogoliubov integro-differential equations can
be time-consuming, especially for nuclei with
non-axial shapes. For that reason a separable form of the pairing interaction
was recently introduced for RHB calculations in spherical and
deformed nuclei \cite{TMR.09a,Niksic10}. The interaction is separable in momentum
space: $\bra{k}V^{^1S_0}\ket{k'} = - G p(k) p(k')$ and,
by assuming a simple Gaussian ansatz $p(k) = e^{-a^2k^2}$, the two parameters $G$ and $a$
were adjusted to reproduce the density dependence of the gap at the Fermi surface
in nuclear matter, calculated with a Gogny force. For the D1S parameterization of
the Gogny force~\cite{BGG.91}, the corresponding parameters of the separable pairing
interaction take the following values:
$G=-728\;{\rm MeV fm}^3$ and $a=0.644\;{\rm fm}$. When
transformed from momentum to coordinate space, the force takes the form:
\begin{equation}
\label{pp-force}
V(\bff{r}_1,\bff{r}_2,\bff{r}_1^\prime,\bff{r}_2^\prime)
=G\delta \left(\bff{R}-\bff{R}^\prime \right)P(\bff{r})P(\bff{r}^\prime)
\frac{1}{2}\left(1-P^\sigma \right),
\end{equation}
where $\bff{R}=\frac{1}{2}\left(\bff{r}_1+\bff{r}_2\right)$ and
$\bff{r}=\bff{r}_1-\bff{r}_2$ denote the center-of-mass and the relative coordinates,
and $P(\bff{r})$ is the Fourier transform of $p(k)$:
\begin{equation}
P(\bff{r}) = \frac{1}{\left(4\pi a^2 \right)^{3/2}}e^{-\bff{r}^2/4a^2} \;.
\end{equation}
The pairing interaction is of finite range and, because of the presence of the factor
$\delta\left(\bff{R}-\bff{R}^\prime \right)$, it preserves translational invariance.
Even though  $\delta\left(\bff{R}-\bff{R}^\prime \right)$ implies that this force is not
completely separable in coordinate space, the corresponding
$pp$ matrix elements can be
represented as a sum of a finite number of separable terms in the basis
of a three-dimensional (3D) harmonic oscillator.
The interaction of Eq.~(\ref{pp-force}) reproduces pairing
properties of spherical and axially deformed nuclei calculated
with the original Gogny force, but  with the important advantage that
the computational cost is greatly reduced.

To describe nuclei with general quadrupole shapes, the
Dirac-Hartree-Bogoliubov equations (\ref{eq:RHB}) are solved by expanding the nucleon
spinors in the basis of a 3D harmonic oscillator in Cartesian coordinates.
In the present calculation of $N=28$ isotones complete convergence is obtained with
$N_f^{max} = 10$ major oscillator shells.
The map of the energy surface as a function of the quadrupole deformation is obtained
by imposing constraints on the axial and triaxial quadrupole moments. The method of
quadratic constraint uses an unrestricted variation of the function
\begin{equation}
\label{eq:quadrupole-constraints1}
\langle \hat{H}\rangle +\sum_{\mu=0,2}{C_{2\mu}
   \left(\langle \hat{Q}_{2\mu}\rangle -q_{2\mu} \right)^2},
\end{equation}
where $\langle \hat{H}\rangle$ is the total energy, and $\langle \hat{Q}_{2\mu}\rangle$
denotes the expectation value of the mass quadrupole operators:
\begin{equation}
\label{eq:quadrupole-constraints2}
\hat{Q}_{20}=2z^2-x^2-y^2 \quad \textrm{and} \quad \hat{Q}_{22}=x^2-y^2 \;.
\end{equation}
$q_{2\mu}$ is the constrained value of the multipole moment, and $C_{2\mu}$
the corresponding stiffness constant \cite{RS.80}.

\subsection{Collective Hamiltonian in Five Dimensions}
The self-consistent solutions of the constrained triaxial RHB equations, i.e. the
single-quasiparticle energies and wave functions for
the entire energy surface as functions of the quadrupole
deformation, provide the microscopic input for the parameters of a
collective Hamiltonian for vibrational and rotational
degrees of freedom  \cite{Ni.09}. The five quadrupole collective
coordinates are parameterized in terms of the two
deformation parameters $\beta$ and $\gamma$, and three Euler angles
$(\phi,\;\theta,\;\psi)\equiv \Omega$, which define the orientation
of the intrinsic principal axes in the laboratory frame.
\begin{equation}
\label{hamiltonian-quant}
\hat{H}_{\rm coll} = \hat{T}_{\textnormal{vib}}+\hat{T}_{\textnormal{rot}}
              +V_{\textnormal{coll}} \; ,
\end{equation}
with the vibrational kinetic energy:
\begin{align}
\hat{T}_{\textnormal{vib}} =&-\frac{\hbar^2}{2\sqrt{wr}}
   \left\{\frac{1}{\beta^4}
   \left[\frac{\partial}{\partial\beta}\sqrt{\frac{r}{w}}\beta^4
   B_{\gamma\gamma} \frac{\partial}{\partial\beta}
   - \frac{\partial}{\partial\beta}\sqrt{\frac{r}{w}}\beta^3
   B_{\beta\gamma}\frac{\partial}{\partial\gamma}
   \right]\right.
   \nonumber \\
   &+\frac{1}{\beta\sin{3\gamma}}\left.\left[
   -\frac{\partial}{\partial\gamma} \sqrt{\frac{r}{w}}\sin{3\gamma}
      B_{\beta \gamma}\frac{\partial}{\partial\beta}
    +\frac{1}{\beta}\frac{\partial}{\partial\gamma} \sqrt{\frac{r}{w}}\sin{3\gamma}
      B_{\beta \beta}\frac{\partial}{\partial\gamma}
   \right]\right\} \; ,
\end{align}
and rotational kinetic energy:
\begin{equation}
\hat{T}_{\textnormal{\textnormal{\textnormal{rot}}}} =
\frac{1}{2}\sum_{k=1}^3{\frac{\hat{J}^2_k}{\mathcal{I}_k}} \; .
\end{equation}
$V_{\textnormal{coll}}$ is the collective potential.
$\hat{J}_k$ denotes the components of the angular momentum in
the body-fixed frame of a nucleus, and the mass parameters
$B_{\beta\beta}$, $B_{\beta\gamma}$, $B_{\gamma\gamma}$, as well as
the moments of inertia $\mathcal{I}_k$, depend on the quadrupole
deformation variables $\beta$ and $\gamma$:
\begin{equation}
\mathcal{I}_k = 4B_k\beta^2\sin^2(\gamma-2k\pi/3) \;.
\end{equation}
Two additional quantities that appear in the expression for the vibrational energy:
$r=B_1B_2B_3$, and $w=B_{\beta\beta}B_{\gamma\gamma}-B_{\beta\gamma}^2 $,
determine the volume element in the collective space.

The dynamics of the collective Hamiltonian is governed by the
seven functions of the intrinsic deformations $\beta$ and $\gamma$:
the collective potential, the three mass parameters:
$B_{\beta\beta}$, $B_{\beta\gamma}$, $B_{\gamma\gamma}$, and the
three moments of inertia $\mathcal{I}_k$. These functions are
determined by the microscopic nuclear energy
density functional and the effective interaction in the $pp$ channel.
The moments of inertia are
calculated from the Inglis-Belyaev formula:
\begin{equation}
\label{Inglis-Belyaev}
\mathcal{I}_k = \sum_{i,j}{\frac{| \langle ij |\hat{J}_k | \Phi \rangle |^2}{E_i+E_j}}\quad k=1,2,3,
\end{equation}
where $k$ denotes the axis of rotation, the summation runs over
proton and neutron quasiparticle states
$|ij\rangle=\beta^\dagger_i\beta^\dagger_j|\Phi\rangle$, and
$|\Phi\rangle$ represents the quasiparticle vacuum. The mass
parameters associated with the two quadrupole collective coordinates
$q_0=\langle\hat{Q}_{20}\rangle$ and $q_2=\langle\hat{Q}_{22}\rangle$
are calculated in the cranking approximation:
\begin{equation}
\label{masspar-B}
B_{\mu\nu}(q_0,q_2)=\frac{\hbar^2}{2}
 \left[\mathcal{M}_{(1)}^{-1} \mathcal{M}_{(3)} \mathcal{M}_{(1)}^{-1}\right]_{\mu\nu}\;,
\end{equation}
where
\begin{equation}
\label{masspar-M}
\mathcal{M}_{(n),\mu\nu}(q_0,q_2)=\sum_{i,j}
 {\frac{\left|\langle\Phi|\hat{Q}_{2\mu}|ij\rangle
 \langle ij |\hat{Q}_{2\nu}|\Phi\rangle\right|}
 {(E_i+E_j)^n}}\;.
\end{equation}
Finally, the potential $V_{\textnormal{coll}}$ in the collective
Hamiltonian Eq.~(\ref{hamiltonian-quant}) is obtained by subtracting
the zero-point energy (ZPE) corrections from the total energy that
corresponds to the solution of constrained RHB equations, at each
point on the triaxial deformation plane  \cite{Ni.09}.

The Hamiltonian Eq.~(\ref{hamiltonian-quant}) describes quadrupole vibrations,
rotations, and the coupling of these collective modes. The corresponding
eigenvalue problem is solved using an expansion of eigenfunctions in terms
of a complete set of basis functions that depend on the deformation variables $\beta$ and
$\gamma$, and the Euler angles $\phi$, $\theta$ and $\psi$ \cite{Ni.09}. The diagonalization
of the Hamiltonian yields the excitation energies and collective wave functions:
\begin{equation}
\label{wave-coll}
\Psi_\alpha^{IM}(\beta,\gamma,\Omega) =
  \sum_{K\in \Delta I}
           {\psi_{\alpha K}^I(\beta,\gamma)\Phi_{MK}^I(\Omega)}.
\end{equation}
The angular part corresponds to linear combinations of Wigner
functions
\begin{equation}
\label{Wigner}
\Phi_{MK}^I(\Omega)=\sqrt{\frac{2I+1}{16\pi^2(1+\delta_{K0})}}
\left[D_{MK}^{I*}(\Omega)+(-1)^ID_{M-K}^{I*}(\Omega) \right] \; ,
\end{equation}
and the summation in Eq. (\ref{wave-coll}) is over the allowed set  of
the $K$ values:
\begin{equation}
\Delta I = \left\{ \begin{array}{lcl}
   0,2,\dots,I \quad &\textnormal{for} \quad  &I\; \textnormal{mod}\; 2 = 0 \\
   2,4,\dots,I-1 \quad &\textnormal{for} \quad   &I\; \textnormal{mod}\; 2 =1\; .
\end{array} \right .
\end{equation}
Using the collective wave functions Eq.~(\ref{wave-coll}),
various observables can be calculated and compared with
experimental results. For instance, the quadrupole E2 reduced
transition probability:
\begin{equation}
\label{BE2}
B(\textnormal{E2};\; \alpha I \to \alpha^\prime I^\prime)=
      \frac{1}{2I+1}|\langle \alpha^\prime I^\prime || \mathcal{\hat{M}}(E2) ||
                                        \alpha I  \rangle|^2 \; ,
\end{equation}
where $\mathcal{\hat{M}}(E2)$ is the electric quadrupole operator,
local in the collective deformation variables.

\section{\label{secIII}Evolution of shapes in the N=28 isotones}

\subsection{\label{ssecI}Quadrupole binding energy maps}

The 3D relativistic Hartree-Bogoliubov model, with the functional DD-PC1 in the
particle-hole channel and a separable pairing force in the particle-particle
channel, enables very efficient constrained self-consistent triaxial calculations
of binding energy maps as functions of quadrupole deformation
in the $\beta - \gamma$ plane. The resulting single-quasiparticle energies
and wave functions provide the microscopic input for
the GCM configuration mixing of angular-momentum
projected triaxial wave functions, or can be used to determine
the parameters of the collective Hamiltonian for vibrations and rotations:
the mass parameters, the moments of inertia, and the collective potential.
The solution of the corresponding eigenvalue problem yields the excitation
spectra and collective wave functions that are used in the calculation of
electromagnetic transition probabilities. This approach is here applied to
the low-energy quadrupole spectra of $N=28$ isotones.

\begin{figure}[htb]
\includegraphics[scale=0.7]{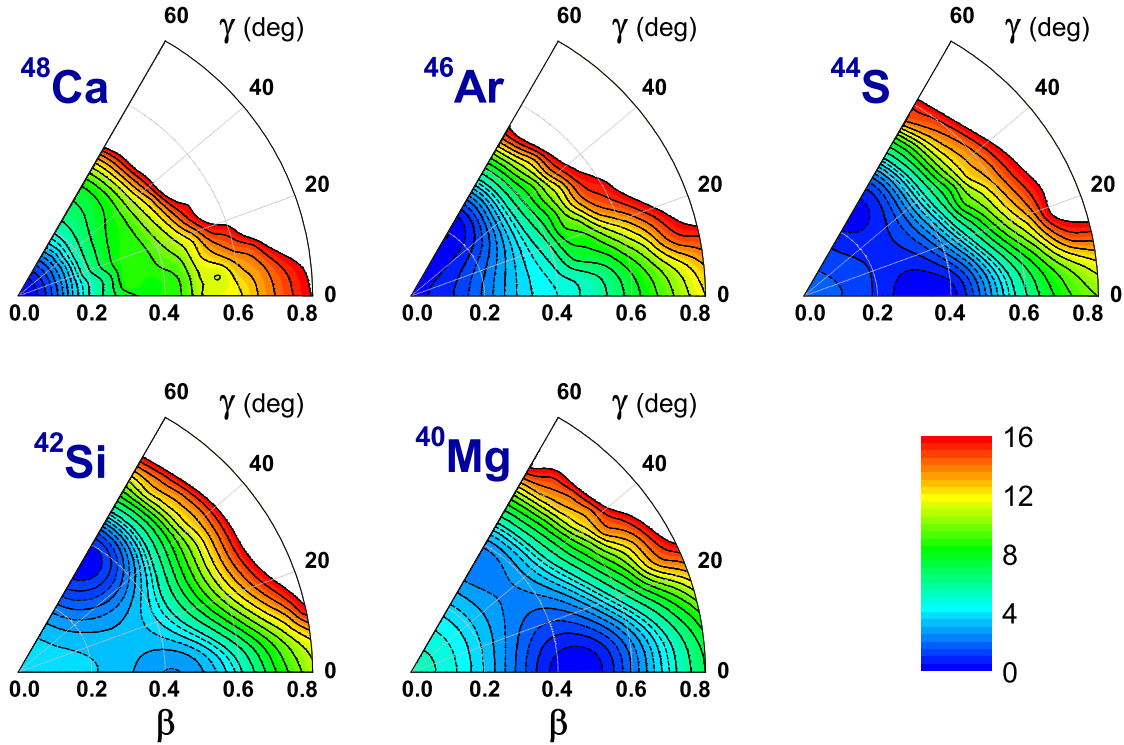}
\caption{\label{fig:PES}(Color online) Self-consistent RHB triaxial
quadrupole constrained energy surfaces of $N=28$ isotones in the $\beta-\gamma$ plane
($0\le \gamma \le 60^0$). For each nucleus energies are normalized with respect to the
binding energy of the global minimum. The contours join points on
the surface with the same energy (in MeV).}
\end{figure}

Figure~\ref{fig:PES} displays the self-consistent
RHB triaxial quadrupole constrained energy surfaces of $N=28$ isotones in the $\beta-\gamma$
plane ($0\le \gamma \le 60^\circ$), calculated using the DD-PC1 energy density functional,
plus the separable pairing force Eq.~(\ref{pp-force}) in the particle-particle channel.
For each nucleus energies are normalized with respect to the
binding energy of the absolute minimum. The contours join points on
the surface with the same energy.

Starting from the spherical doubly-magic $^{48}$Ca, we consider the even-even
$N=28$ isotones obtained by successive removals of proton pairs. The binding
energy maps display a rich variety of rapidly evolving shapes, and clearly
demonstrate the fragility of the $N=28$ shell. By removing a pair of
protons from $^{48}$Ca, the energy surface of the corresponding isotone $^{46}$Ar
becomes soft both in $\beta$ and $\gamma$, with a shallow extended minimum
along the oblate axis. Only four protons away from the doubly magic $^{48}$Ca,
DD-PC1 predicts a coexistence of prolate and oblate  minima
at $(\beta,\gamma)=(0.34, 0^\circ)$ and $(0.27, 60^\circ)$, respectively, in $^{44}$S.
The two minima are separated by a rather low barrier of less than 1 MeV and,
therefore, one expects to find pronounced mixing of prolate and oblate configurations
in the low-energy collective states of this nucleus.
For $^{42}$Si the binding energy displays a deep oblate minimum at
$(\beta,\gamma)=(0.35, 60^\circ)$, whereas a secondary, prolate minimum is calculated
$\sim2.5$~MeV higher. Finally, with another proton pair removed, the
very neutron-rich nucleus $^{40}$Mg shows a deep prolate minimum
at $(\beta,\gamma)=(0.45, 0^\circ)$.

We note that similar binding energy surfaces
were also obtained in recent studies~\cite{Delaroche10,CEA} based on the
self-consistent Hartree-Fock-Bogoliubov (HFB) model, using the finite-range
and density-dependent Gogny D1S interaction. On the mean-field level
the only qualitative difference is found for $^{40}$Mg. For this nucleus
the present calculation predicts a saddle point on the oblate axis, whereas
a secondary local oblate minimum is obtained in
the HFB calculation with the Gogny force.

\begin{figure}[htb]
\includegraphics[scale=0.4]{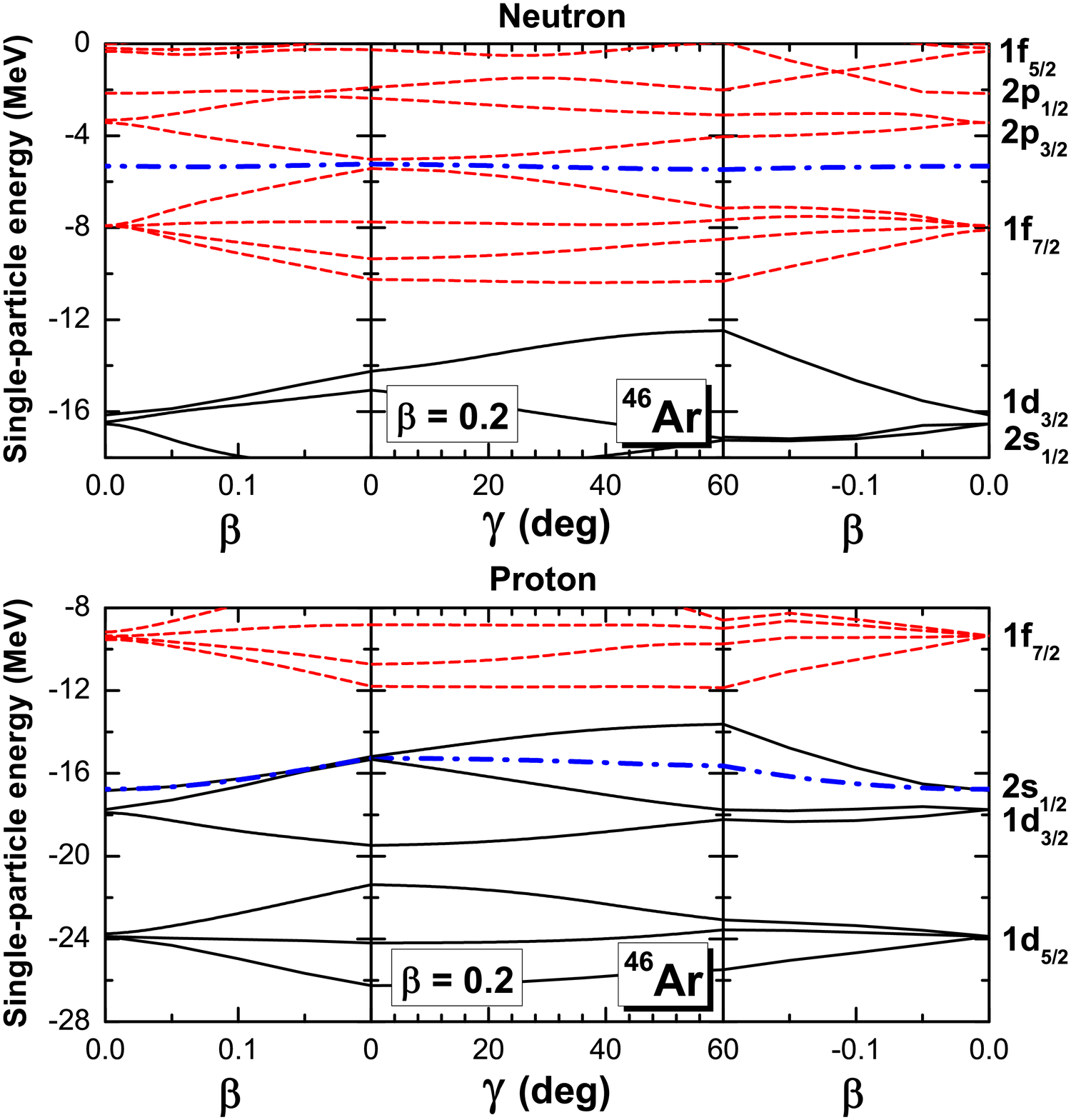}
\caption{\label{fig:Esp-bg-Ar46}(Color online) Single-neutron and
single-proton energy levels of $^{46}$Ar, as functions of the
deformation parameters along closed paths in the $\beta - \gamma$ plane.
Solid (black) curves correspond to levels with positive parity, and
(red) dashed curves denote levels with negative parity.
The dot-dashed (blue) curves corresponds to the Fermi levels.
The panels on the left and right display prolate ($\gamma =0^\circ$) and oblate
 ($\gamma =60^\circ$) axially-symmetric single-particle levels, respectively.
 In the middle panel of each figure the neutron and proton levels are
 plotted as functions of $\gamma$, for a fixed value of the axial deformation
 $|\beta|$ at the approximate position of the mean-field minimum.}
\end{figure}

\begin{figure}[htb]
\includegraphics[scale=0.4]{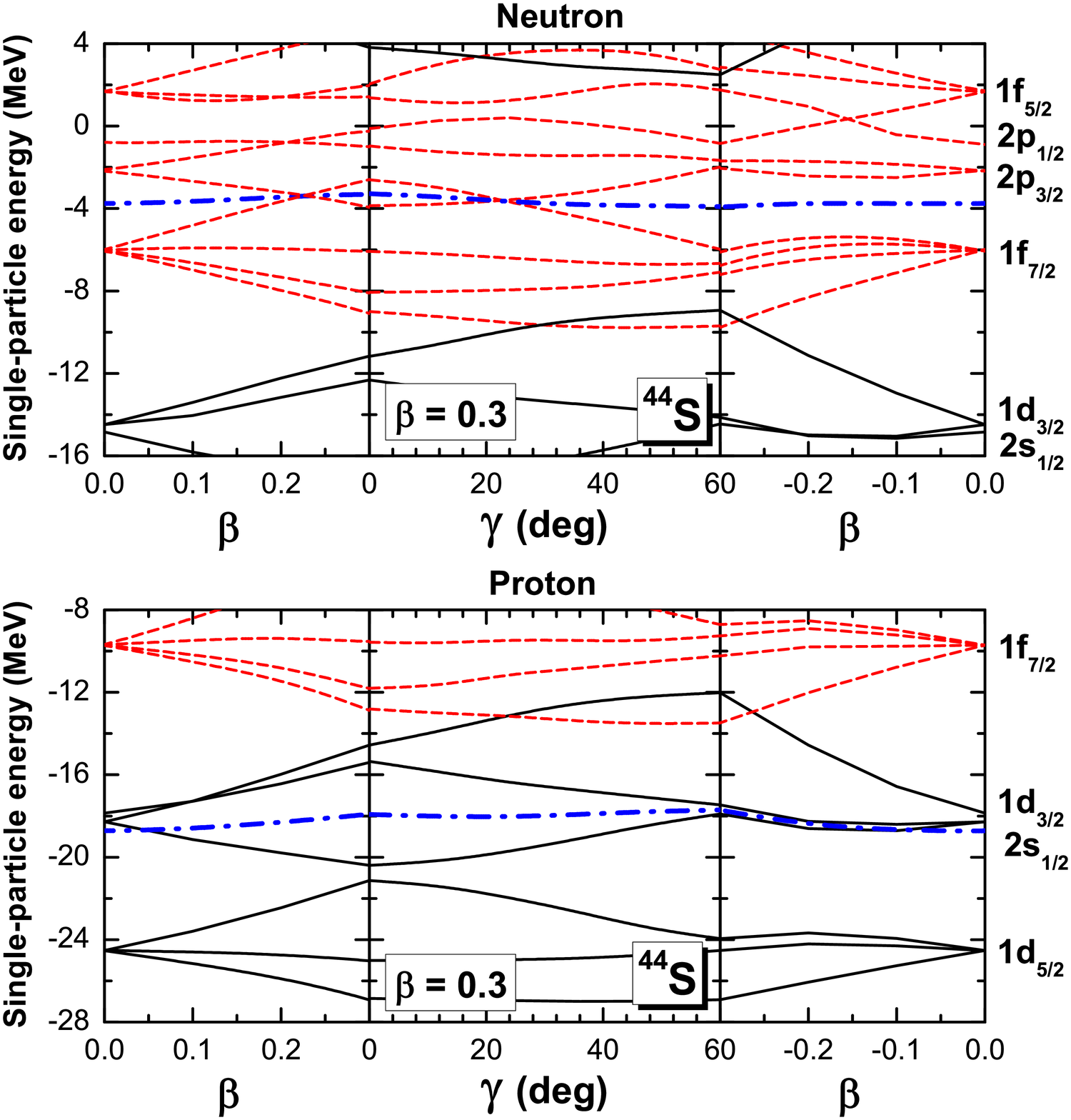}
\caption{\label{fig:Esp-bg-S44}(Color online) Same as
described in the caption to Fig. \ref{fig:Esp-bg-Ar46}
but for the nucleus $^{44}$S.}
\label{44S}
\end{figure}

\begin{figure}[htb]
\includegraphics[scale=0.4]{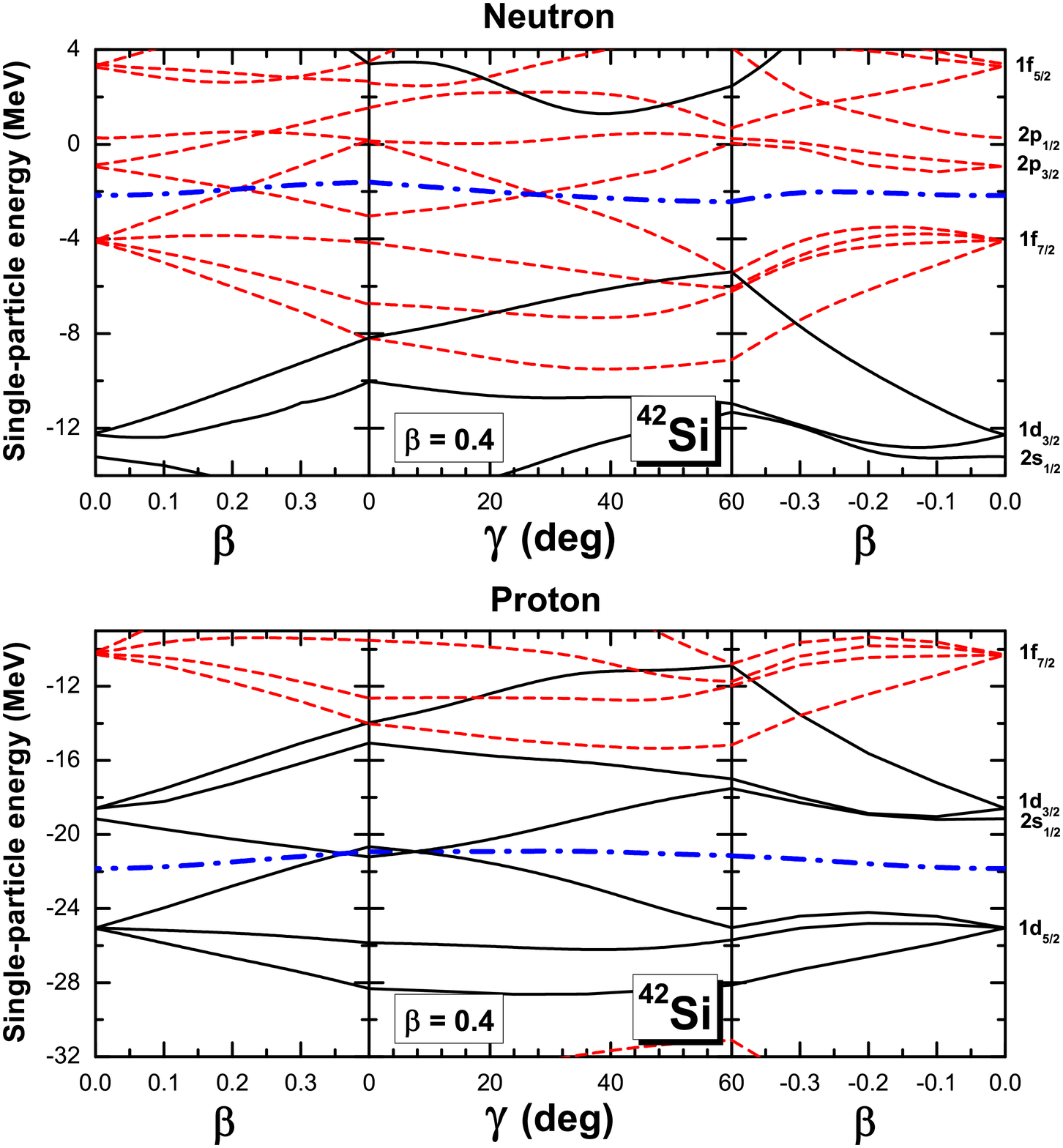}
\caption{\label{fig:Esp-bg-Si42}(Color online) Same as described in the caption to
Fig. \ref{fig:Esp-bg-Ar46}
but for the nucleus $^{42}$Si.}
\end{figure}

\begin{figure}[htb]
\includegraphics[scale=0.4]{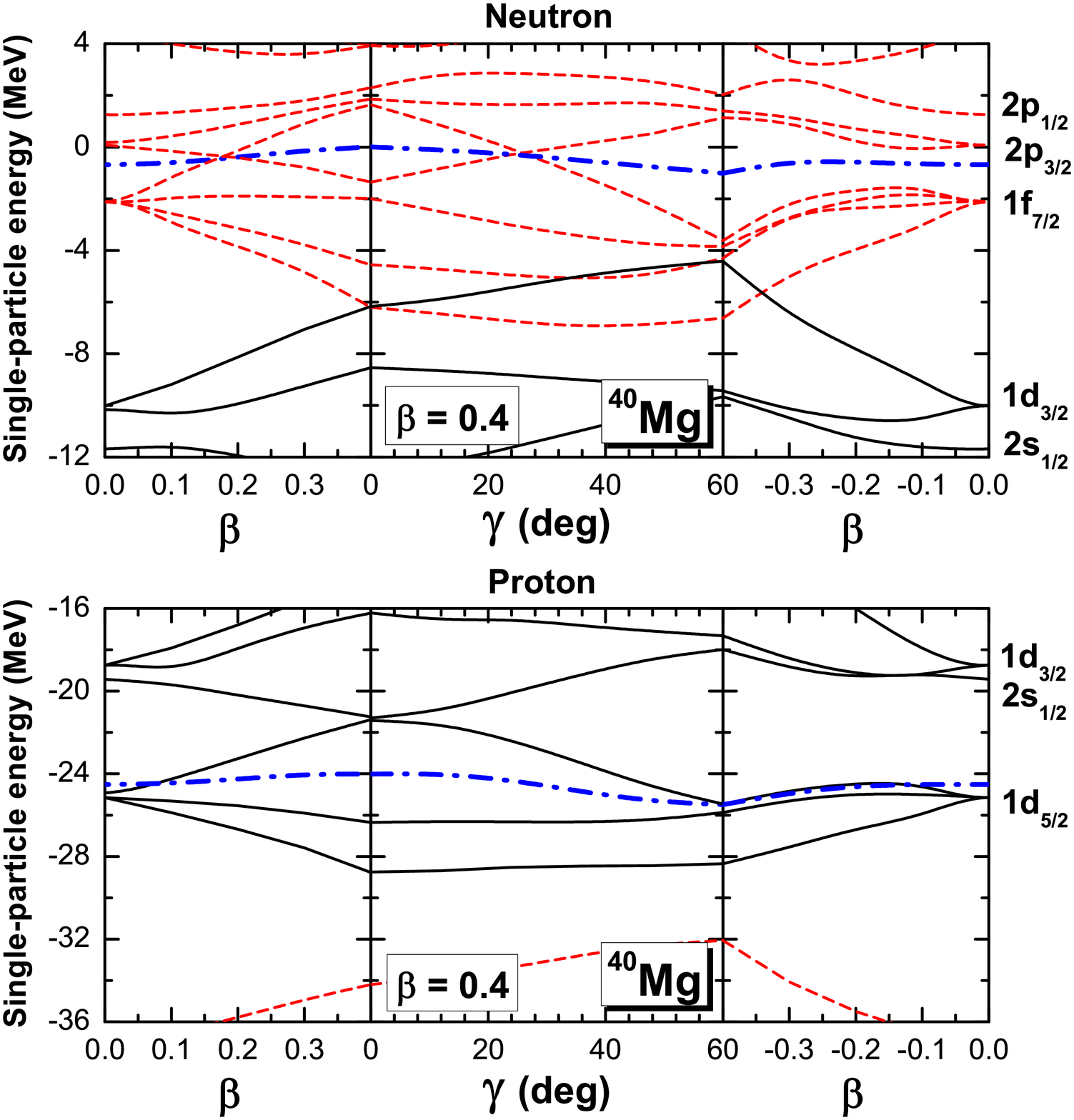}
\caption{\label{fig:Esp-bg-Mg40}(Color online) Same as described in the caption to
Fig. \ref{fig:Esp-bg-Ar46}
but for the nucleus $^{40}$Mg.}
\end{figure}

The variation of mean-field shapes in an isotopic, or isotonic, chain
is governed by the evolution of the underlying shell structure of single-nucleon
orbitals. The formation of deformed minima, in particular, can be related to the
occurrence of gaps or regions of low single-particle level density around the
Fermi surface. In Figs.~\ref{fig:Esp-bg-Ar46} -- \ref{fig:Esp-bg-Mg40} we plot the
neutron and proton single-particle energy levels in the canonical basis for
$^{46}$Ar, $^{44}$S, $^{42}$Si, and $^{40}$Mg, respectively.
Solid (black) curves correspond to levels with positive parity, and
(red) dashed curves denote levels with negative parity.
The dot-dashed (blue) curves correspond to the Fermi levels.
The neutron and proton levels are plotted as functions of the
deformation parameters along closed paths in the $\beta - \gamma$ plane.
The panels on the left and right display prolate ($\gamma =0^\circ$) and oblate
 ($\gamma =60^\circ$) axially-symmetric single-particle levels, respectively.
 In the middle panel of each figure the neutron and proton levels are
 plotted as functions of $\gamma$, for a fixed value of the axial deformation
 $|\beta|$ at the approximate position of the mean-field minima:
 $|\beta|$ = 0.2 for $^{46}$Ar, $|\beta|$ = 0.3 for $^{44}$S,
and $|\beta|$ = 0.4 for $^{42}$Si and $^{40}$Mg. In this way, starting from the spherical configuration,
we follow the single-nucleon levels on a path along the prolate axis up to the approximate position
of the minimum (left panel), then for this fixed value of $|\beta|$ the path from
$\gamma =0^\circ$ to $\gamma =60^\circ$ (middle panel) and, finally, back to the spherical
configuration along the oblate axis (right panel). Negative values of beta denote
axial deformations with $\gamma =60^\circ$, that is, points along the oblate axis.

Figures~\ref{fig:Esp-bg-Ar46} -- \ref{fig:Esp-bg-Mg40} elucidate the principal
characteristics of structural changes in neutron-rich $N=28$ nuclei: the near degeneracy
of the $d_{3/2}$ and $s_{1/2}$ proton orbitals, and the reduction of the size of the
$N=28$ shell gap \cite{OSor.10}.
Between the doubly magic $^{48}$Ca and $^{46}$Ar the spherical gap $N=28$
decreases from 4.73 MeV to 4.48 MeV (cf. Table~\ref{shell}), in excellent
agreement with data: from 4.80 MeV in $^{48}$Ca to 4.47 MeV in
$^{46}$Ar \cite{Gaudefroy06,Gaudefroy07}. Nevertheless, the gap between
occupied and unoccupied neutron levels in $^{46}$Ar is still largest at the
spherical configuration, as shown in the upper panel of Fig.~\ref{fig:Esp-bg-Ar46}.
We note, in particular, the agreement of the calculated energies of spherical neutron
states with experimental single-neutron energies \cite{Gaudefroy06}.
For the proton states shown in the lower panel, the largest gap is found
at $|\beta|$ = 0.2 and $\gamma =60^\circ$, that is, on the oblate axis. The
competition between the spherical configuration favored by neutron states and
the oblate shape favored by proton states, leads to the shallow extended oblate
minimum shown in Fig.~\ref{fig:PES}.
Two protons less, and the spherical $N=28$ gap is reduced by another
620 keV to 3.86 MeV in $^{44}$S. The largest gap between neutron states
is not the spherical one like in $^{46}$Ar, however, but at the oblate deformation
$|\beta| \approx$ 0.3 and
$\gamma =60^\circ$ (upper panel of Fig.~\ref{fig:Esp-bg-S44}). The removal of two
protons lowers the energy of the corresponding Fermi level, and for $^{44}$S the largest gap
is found on the prolate axis (lower panel of Fig.~\ref{fig:Esp-bg-S44}). The formation
of the oblate neutron and prolate proton gaps is at the origin of the
coexistence of deformed shapes in $^{44}$S (cf. Fig.~\ref{fig:PES}).
In $^{42}$Si both neutron and proton gaps are on the oblate axis resulting in
the pronounced oblate minimum at $|\beta| \approx$ 0.35. Finally, the deep prolate
minimum at $\beta \approx$ 0.35 in $^{40}$Mg arises because of the neutron gap and,
especially pronounced, proton gap on the prolate axis. We note that the largest
neutron gap for this nucleus is still on the oblate side but, because the protons
strongly favor the prolate configuration, it produces only a saddle point on the
oblate axis, as shown in Fig.~\ref{fig:PES}.
\begin {table}[h]
\begin {center}
\caption{The DD-PC1 RHB theoretical neutron $N=28$ spherical energy gaps, and the
corresponding values of the axial deformation for the minima of the quadrupole
binding energy maps of $^{48}$Ca, $^{46}$Ar, $^{44}$S, $^{42}$Si, and $^{40}$Mg.
Negative values of $\beta$ denote oblate shapes.}
\bigskip
\begin {tabular}{c|cc}
\hline
\hline
 & $\Delta^{\rm sph.}_{N=28}$  & \ \ $\beta_{\rm min}$\ \   \\
\hline
$^{48}$Ca  & 4.73  &  0.00  \\
$^{46}$Ar  & 4.48  & -0.19  \\
$^{44}$S   & 3.86  &  0.34  \\
$^{42}$Si  & 3.13  & -0.35  \\
$^{40}$Mg  & 2.03  &  0.45  \\
\hline
\end{tabular}
\label{shell}
\end{center}
\end{table}

The erosion of the spherical $N=28$ shell is also shown in Table~\ref{shell},
where we include the DD-PC1 RHB theoretical neutron $N=28$ spherical energy gaps, and the
corresponding values of the axial deformation for the minima of the quadrupole
binding energy maps of $^{48}$Ca, $^{46}$Ar, $^{44}$S, $^{42}$Si, and $^{40}$Mg.
Both experiment and theory point toward a strong reduction of the $N=28$
gap as more protons are removed and, thus, the isotones become more neutron-rich.
N=28 is the first ``magic" number produced by the spin-orbit part of the single-nucleon potential
and, therefore, a relativistic mean-field model automatically reproduces the N=28 gap because
it naturally includes the spin-orbit interaction and the correct isospin dependence of this term,
as it was already shown in the axial RHB calculation of neutron-rich N=28 nuclei \cite{Lalazissis99}.
Experimentally, indirect evidence of the erosion of the gap has been obtained
by following the evolution of excitation energies of the $2^+_1$ state and
the E2 transitions in $N=28$ isotones and
neighboring nuclei~\cite{Sorlin93,Scheit96,Glasmacher97,Gaudefroy06,Force10}.
The experimental results can be reproduced by both mean-field~\cite{Lalazissis99,Peru00} and
shell model~\cite{Nowacki09} calculations. As shown in
Table~\ref{shell}, the DD-PC1 RHB calculation predicts a reduction of the spherical $N=28$ shell gap
from 4.73~MeV in the doubly-magic nucleus $^{48}$Ca to 2.03~MeV in the well-deformed $^{40}$Mg.
We note that the theoretical values of the spherical shell gap for $^{48}$Ca and $^{46}$Ar are very close
to data: 4.80 MeV in $^{49}$Ca, and 4.47 MeV in $^{47}$Ar,
obtained by neutron stripping reactions~\cite{Gaudefroy06,Gaudefroy07}.

\begin{figure}[htb]
\includegraphics[scale=0.45]{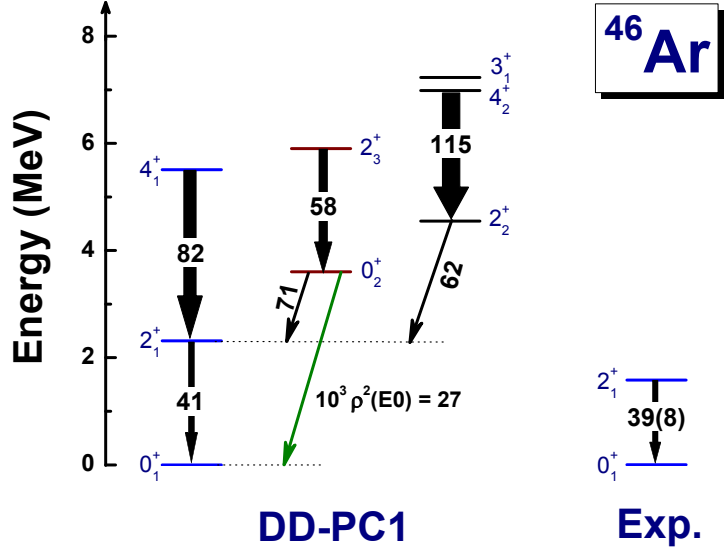}
\caption{(Color online)
The spectrum of $^{46}$Ar calculated with the
DD-PC1 relativistic density functional  (left),
compared to data \cite{NNDC} (right) for the excitation energy of $2^+_1$,
and the reduced electric quadrupole transition B(E2) (in units of $e^2{\rm fm}^4$).
The prediction for the electric monopole transition strength $\rho^2(E0; 0^+_2\rightarrow 0^+_1)$
is also included in the theoretical spectrum.}
\label{fig:spec-Ar46}
\end{figure}
\begin{figure}[htb]
\includegraphics[scale=0.45]{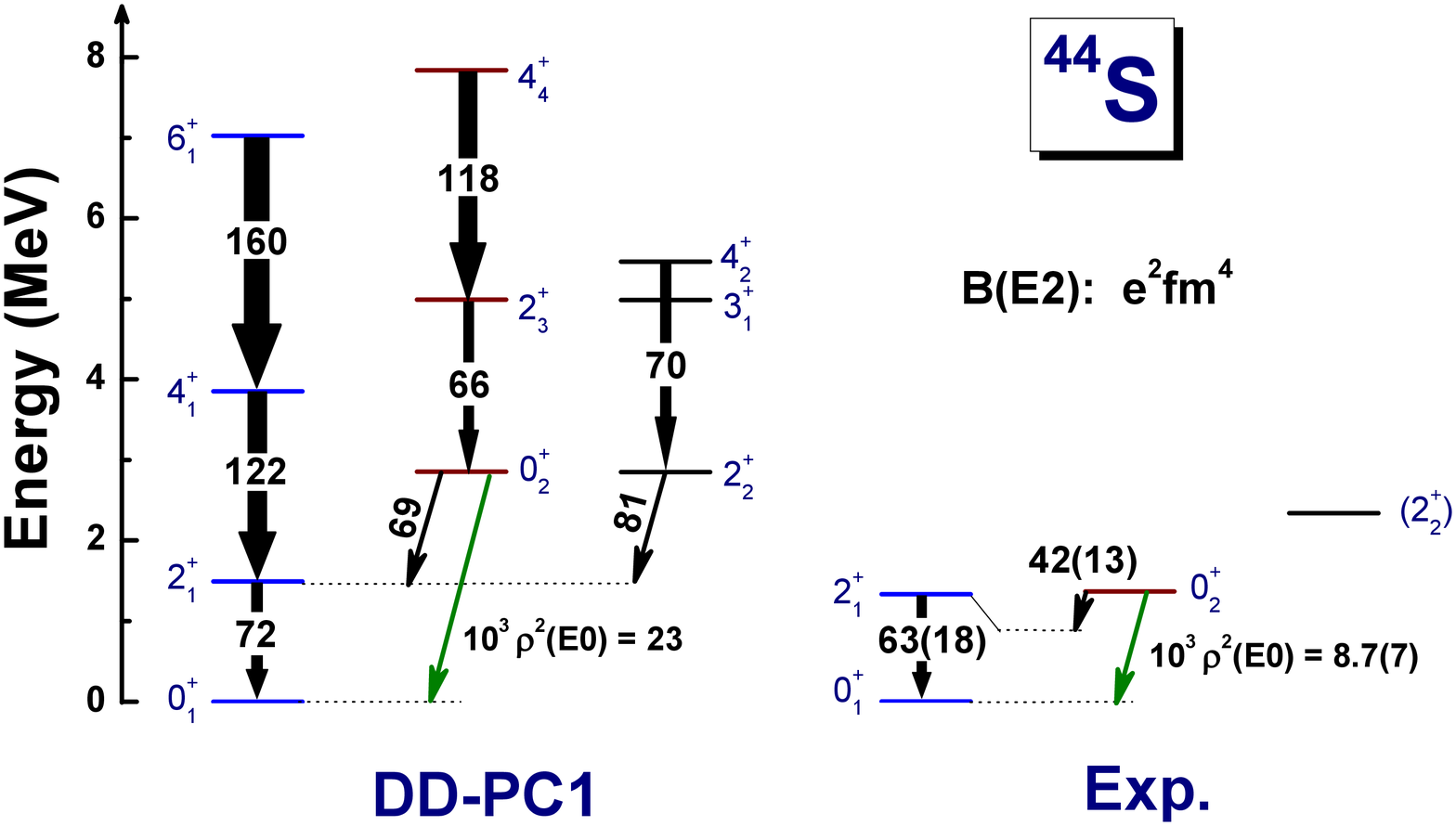}
\caption{(Color online)
Same as described in the caption to Fig. \ref{fig:spec-Ar46} but for the nucleus $^{44}$S.
The data are from Refs.~\cite{Glasmacher97,Force10}.}
\label{fig:spec-S44}
\end{figure}
\begin{figure}[htb]
\includegraphics[scale=0.45]{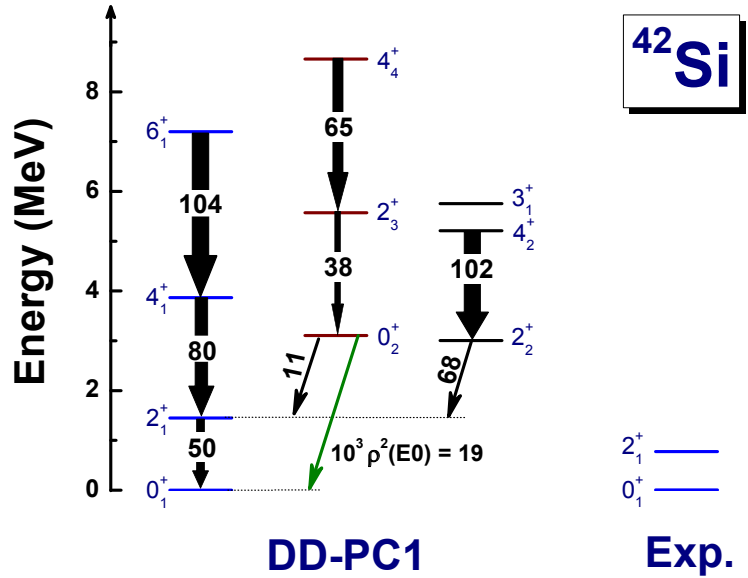}
\caption{(Color online)
Same as described in the caption to Fig. \ref{fig:spec-Ar46} but for the nucleus $^{42}$Si.
The data are from Ref.~\cite{Bastin07}.}
\label{fig:spec-Si42}
\end{figure}

\subsection{\label{ssecIII}Low-energy collective spectra}

Starting from constrained self-consistent solutions of the RHB equations,
that is, using single-quasiparticle energies and wave functions that correspond to
each point on the energy surfaces shown in Fig.~\ref{fig:PES}, the
parameters that determine the collective Hamiltonian: the mass parameters
$B_{\beta\beta}$, $B_{\beta\gamma}$, $B_{\gamma\gamma}$, three
moments of inertia $\mathcal{I}_k$, as well as the zero-point energy
corrections, are calculated as functions of the quadrupole deformations $\beta$
and  $\gamma$. The diagonalization of the resulting Hamiltonian yields the
excitation energies and reduced transition probabilities.
In Figs.~\ref{fig:spec-Ar46} -- \ref{fig:spec-Si42} we display
the spectra of $^{46}$Ar, $^{44}$S, and $^{42}$Si calculated with the
DD-PC1 relativistic density functional plus the separable pairing force Eq.~(\ref{pp-force}),
in comparison to available data for the excitation energies, reduced electric quadrupole
transition probabilities B(E2) (in units of $e^2{\rm fm}^4$), and the electric monopole transition
strength $\rho^2(E0; 0^+_2\rightarrow 0^+_1)$. We emphasize that this
calculation is completely parameter-free, that is, by using the self-consistent
solutions of the RHB single-nucleon equations, physical observables, such as transition
probabilities and spectroscopic quadrupole moments, are calculated in
the full configuration space and there is no need for effective charges.
Using the bare value of the proton charge in the electric
quadrupole operator, the transition probabilities between eigenstates of the
collective Hamiltonian can directly be compared to data.

\begin{figure}[htb]
\includegraphics[scale=0.45]{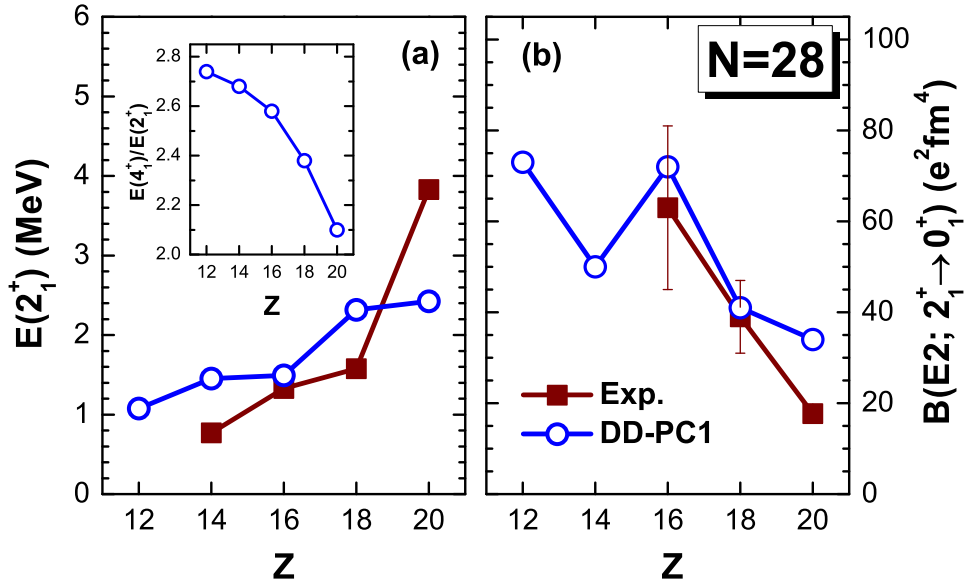}
\caption{\label{fig:obs}(Color online) Evolution of the characteristic
observables $E(2^+_1)$ and $B(E2; 2^+_1\rightarrow 0^+_1 )$ (in ${\rm e}^2 {\rm fm}^4$)
with proton number in $N=28$ isotones. The ratio between the excitation energies of the
first $4^+$ and $2^+$ states is also displayed in the inset. The microscopic values calculated
with the energy density functional DD-PC1 are shown in comparison with available data.}
\end{figure}

Before considering the excitation spectra of individual nuclei and, in particular, shape coexistence
in $^{44}$S, in Fig.~\ref{fig:obs} we illustrate the evolution with proton number of characteristic
collective observables: the excitation energy of the first $2^+$
state, the ratio $E(4^+_1)/E(2^+_1)$, and the reduced transition probability B(E2; $2^+_1 \to 0^+_1$).
The rapid decrease of the ratio $E(4^+_1)/E(2^+_1)$ from $\approx 2.8$ in $^{40}$Mg to $\approx 2.1$
in $^{48}$Ca is characteristic for a transition from a deformed rotational nucleus to a spherical vibrator.
Note, however, that even in the case of $^{40}$Mg the value of $E(4^+_1)/E(2^+_1)$ is considerably
below the rigid-rotor limit of 3.3. The excitation energy of the first excited state
$2^+_1$ can directly be compared to data. The calculated
$E(2^+_1)$ increases with proton number toward the doubly magic $^{48}$Ca, but the
predicted rise in energy is not as sharp as in experiment. In fact, one expects that in deformed nuclei,
e.g $^{42}$Si, the calculated $E(2^+_1)$ is above the experimental excitation energy, because of
the well-known fact that the Inglis-Belyaev formula Eq.~(\ref{Inglis-Belyaev}) predicts
effective moments of inertia that are smaller than empirical values. The moments of
inertia can generally be improved by including the Thouless-Valatin (TV) dynamical rearrangement
contributions \cite{Delaroche10}, but the calculation of the
TV moments of inertia~\cite{Thouless62} has not yet been implemented in the
collective Hamiltonian used in the present calculation. The panel on the right of Fig.~\ref{fig:obs}
displays the evolution with proton number of another
characteristic collective observable: $B(E2; 2^+_1 \to 0^+_1)$ (in $e^2{\rm fm}^4$).
The calculation reproduces the empirical decrease of $B(E2; 2^+_1 \to 0^+_1)$
with proton number and, in particular, we notice the excellent agreement
between the parameter-free theoretical predictions and data for $^{44}$S and $^{46}$Ar.

Figure~\ref{fig:spec-Ar46} displays the low-energy spectrum of $^{46}$Ar. The
excitation energy  $E(2^+_1)$ is calculated considerably above the experimental state,
whereas the $B(E2; 2^+_1 \to 0^+_1)$ reproduces the experimental value. In the present
analysis we particularly focus on $^{44}$S, for which data that indicate shape coexistence
were reported recently \cite{Force10}. Already the data on the low energy of the first $2^+$ state
and the enhanced $B(E2; 2^+_1 \to 0^+_1)$ of 63(18) $e^2{\rm fm}^4$ \cite{Glasmacher97} pointed
towards a possible deformation of the ground state of $^{44}$S. More recently, the structure of this
nucleus was studied by using delayed $\gamma$ and electron spectroscopy, and new data
were reported for the reduced transition
probability $B(E2; 2^+_1 \to 0^+_2) = 8.4 (26)$ $e^2{\rm fm}^4$, and the monopole
strength $\rho^2(E0; 0^+_2 \to 0^+_1) = 8.7 (7) \times 10^{-3}$ \cite{Force10}. From a
comparison to shell model calculations, a prolate-spherical shape coexistence was inferred, and a
two-level mixing model was used to extract a weak mixing between the two configurations.
The spectrum of $^{44}$S calculated in this work is compared to available data in Fig.~\ref{fig:spec-S44}.
The model nicely reproduces both the excitation energy and the reduced transition probability
$B(E2; 2^+_1 \to 0^+_1)$ for the first excited state $2^+_1$, and the theoretical value for
$B(E2; 0^+_2 \to 2^+_1)$ is also in good agreement with data. The experimental ratio
$B(E2; 2^+_1 \to 0^+_1) / B(E2; 2^+_1 \to 0^+_2)$ is 7.5, and the
calculated value is 5.2. The excitation energy of
the state $0^+_2$, however, is calculated much higher than the experimental counterpart.
Together with the fact that the calculated monopole transition strength
$\rho^2(E0; 0^+_2\rightarrow 0^+_1) (\times 10^3)$ = 23 is larger than the
corresponding experimental value of 8.7(7), this result indicates that there is more
mixing between the theoretical states $0^+_1$ and $0^+_2$ than what can be inferred from
the data.

The low-lying $0^+_2$ state with the excitation energy $1.365$ MeV, the rather weak interband
transition probability B(E2; $2^+_1 \to 0^+_2$)=8.4(26) $e^2{\rm fm}^4$, and the
monopole strength $\rho^2(E0; 0^+_2\to 0^+_1)=8.7(7)\times10^{-3}$ have been regarded
as fingerprints of shape coexistence in $^{44}$S~\cite{Force10}. One reason for the more
pronounced mixing between the calculated $0^+_1$ and $0^+_2$ in this work and, consequently, the
higher excitation energy of $0^+_2$, could be the particular choice of the energy density functional
and/or the treatment of pairing correlations~\cite{Li11}. The predicted barrier between the
prolate and oblate minima (cf. Fig.~\ref{fig:PES}) could, in fact, be too low. Another reason
for the high excitation energy of $0^+_2$ could
be the approximation used in the calculation of mass parameters
(vibrational inertial functions). In the current version of the
model the mass parameters are determined by using the cranking approximation
Eqs.~(\ref{masspar-B}) and (\ref{masspar-M}), in which the time-odd components
(the so-called Thouless-Valatin dynamical rearrangement contributions) are omitted.
Recently an efficient microscopic derivation of the five-dimensional
quadrupole collective Hamiltonian has been developed, based on the adiabatic self-consistent
collective coordinate method~\cite{Hinohara10}. In this model the moments of inertia and mass parameters
are determined from local normal modes built on constrained Hartree-Fock-Bogoliubov states,
and the TV dynamical rearrangement contributions are treated self-consistently.
For the illustrative case of $^{68}$Se, it has been shown that the self-consistent inclusion of
the time-odd components of the mean-field can lead to an increase of
the mass parameters by $30\%\sim 200\%$, depending on the deformation.
In fact, in the present calculation an enhancement of the cranking masses by a factor $\sim2$
brings the calculated excitation energies, and also the monopole strength $\rho^2(E0; 0^+_2\to 0^+_1)$,
in very close agreement with the experimental spectrum.

In Table~\ref{speS44} we compare the experimental excitation energies of the states $2^+_1$, $0^+_2$,
and $2^+_2$, the reduced transition probabilities $B(E2; 2^+_1\rightarrow 0^+_1)$ ($e^2{\rm fm}^4$),
$B(E2; 2^+_1\rightarrow 0^+_2)$, and the monopole strength $\rho^2(E0; 0^+_2\to 0^+_1)\times10^{3}$
in $^{44}$S, to the results of the present work, the five-dimensional GCM(GOA) calculation with the Gogny
D1S interaction~\cite{CEA}, the angular-momentum projected GCM calculation restricted to axial shapes
(AMP GCM) with the Gogny D1S interaction~\cite{Guzman02} ,
and to shell-model calculations~\cite{Force10}. One might notice that all three models based on constrained
self-consistent mean-field calculations of the binding energy maps (curves in the case of axially-symmetric
AMP GCM), reproduce the data with similar accuracy. It is interesting that only the axially-symmetric
calculation reproduces the very low excitation energy of the state $0^+_2$, whereas the result of
the five-dimensional GCM(GOA) calculation, although it was also based on the Gogny D1S interaction,
is even above the energy obtained with DD-PC1.
Table~\ref{speS44} shows that the best overall agreement with data is obtained in the shell-model (SM)
calculation of Ref.~\cite{Force10}, using the effective interaction SDPF-U \cite{Nowacki09} for $0\hbar\omega$ SM
calculations in the $sd-pf$ valence space, and with a particular choice of the proton and neutron effective
charges.

\begin {table}[htb]
\begin {center}
\caption{Excitation energies (in MeV) of the states $2^+_1$, $0^+_2$, and $2^+_2$,
$B(E2; 2^+_1\rightarrow 0^+_1)$ ($e^2{\rm fm}^4$),
$B(E2; 2^+_1\rightarrow 0^+_2)$, and the monopole strength $\rho^2(E0; 0^+_2\to 0^+_1)\times10^{3}$
in $^{44}$S. The experimental values~\cite{Sohler02,Force10} are compared to the
results of the present work,
the five-dimensional GCM(GOA) calculation with the Gogny D1S interaction~\cite{CEA}, the
angular-momentum projected GCM calculation restricted to axial shapes (AMPGCM)
with the Gogny D1S interaction~\cite{Guzman02} ,
and to shell-model calculations~\cite{Force10}.}
\bigskip
\begin {tabular}{cccccc}
\hline
\hline
 & Experiment & This work  & GCM(GOA) \cite{CEA} & AMPGCM  \cite{Guzman02}& Shell Model  \cite{Force10} \\
\hline
$E(2^+_1)$                       & 1.329(1)  & 1.491 & 1.267 &  1.410  &  1.172 \\
$E(0^+_2)$                       & 1.365(1)  & 2.852 & 3.611 &  1.070  &  1.137 \\
$E(2^+_2)$                       & 2.335(39) & 2.851 & 2.557 &  1.830  &  2.140 \\
$B(E2; 2^+_1\rightarrow 0^+_1)$  & 63(18)    & 72    & 105   &  75     &  75    \\
$B(E2; 2^+_1\rightarrow 0^+_2)$  & 8.4(2.6)   & 14    & 6.3   &  -      &  19    \\
$\rho^2(E0; 0^+_2\to 0^+_1)(\times10^{3})$ & 8.7(7) & 23 & 5.4 &  -    &  -     \\
\hline
\end{tabular}
\label{speS44}
\end{center}
\end{table}


Based on the data included in Table~\ref{speS44} and on the SM calculation with the SDPF-U effective
interaction, in Ref.~\cite{Force10} it was deduced that $^{44}$S exhibits a shape coexistence between
a prolate ground state ($\beta \approx 0.25)$ and a rather spherical $0^+_2$ state. The sequence of
ground-state band states  $0^+_1$, $2^+_1$, $4^+_2$, and $6^+_2$, is connected by strong E2
transitions, and the excited states are characterized by the intrinsic quadrupole moment $Q_0 \approx 60$
$e~{\rm fm}^2$. This sequence was interpreted as a rotational band of an axially deformed prolate shape with
$\beta \approx 0.25$. The calculated $2^+_2$ state has a smaller quadrupole moment $Q_0 = -0.3$
$e~{\rm fm}^2$, compatible with a spherical shape, and is connected by a strong E2 transition to the
$0^+_2$ state. These SM results, therefore, indicate a prolate-spherical shape coexistence
in $^{44}$S \cite{Force10}.

\begin{figure}[htb]
\includegraphics[scale=0.75]{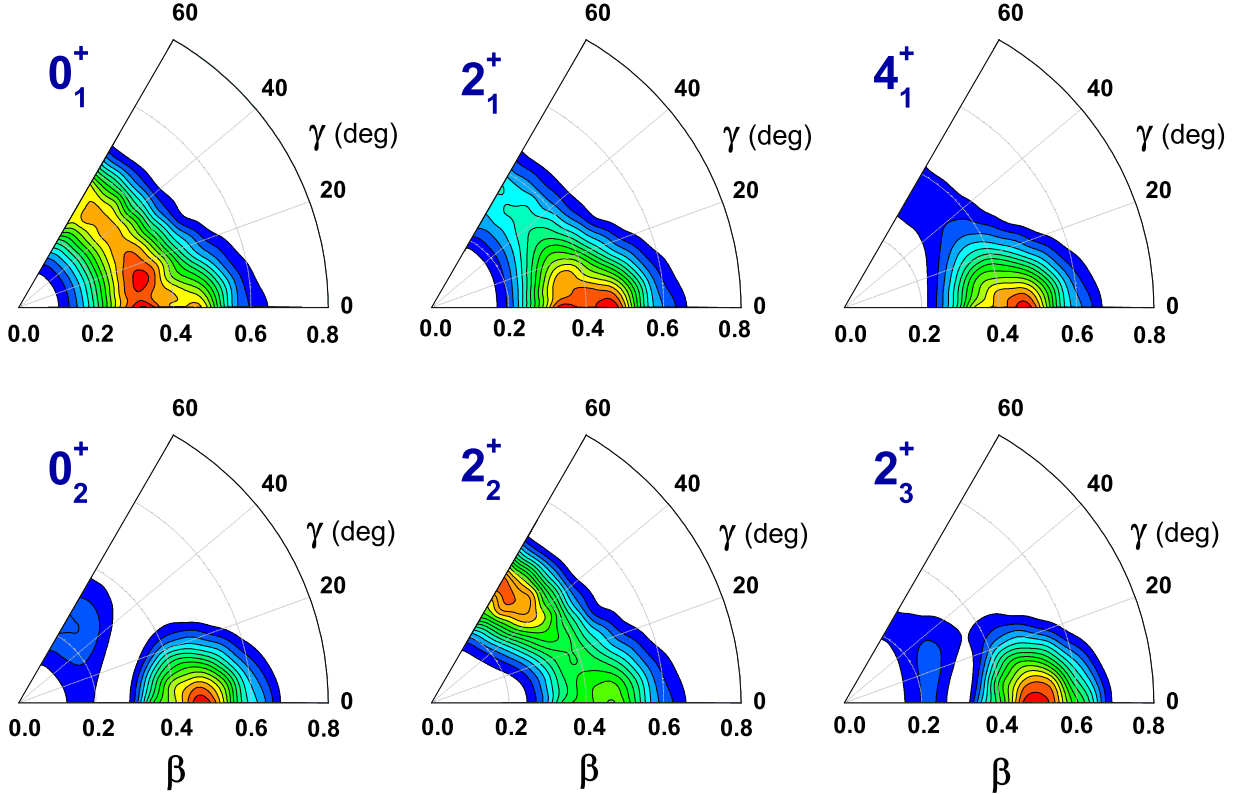}
\caption{\label{wavef-S44}  (Color online) Probability distribution Eq.~(\ref{eq:density})
in the $\beta - \gamma$ plane for the lowest collective states of $^{44}$S,
predicted by DD-PC1 energy density functional. }
\end{figure}

To analyze configuration mixing in the low-energy spectrum based on the functional DD-PC1, in
Fig.~\ref{wavef-S44} we plot the probability density distributions for the three lowest states of the
ground-state band: $0^+_1$, $2^+_1$, and $4^+_1$, the state $0^+_2$, and the two states
$2^+_2$ and $2^+_3$. For a given collective state Eq.~(\ref{wave-coll}), the probability density
distribution in the $(\beta,\gamma)$ plane is defined by:
\begin{equation}
  \label{eq:density}
  \rho_{I\alpha}(\beta,\gamma)=\sum\limits_{K\in\Delta I}|\psi^I_{\alpha K}(\beta,\gamma)|^2
  \beta^3,
\end{equation}
with the normalization:
\begin{equation}
  \int^\infty_0 \beta {\rm d}\beta \int^{2\pi}_0 \rho_{I\alpha}(\beta,\gamma)\ |\sin3\gamma| ~{\rm d}\gamma =1.
\end{equation}
The probability distribution of the ground state $0^+_1$ indicates a deformation
$|\beta| \geq 0.3$, extended in the $\gamma$ direction from the prolate configuration at $\gamma = 0$ to
the oblate configuration at $\gamma = 60^\circ$. The average deformation is
$(\langle\beta\rangle,\langle\gamma\rangle) = (0.32,26^\circ)$, and the $\gamma$-softness
reflects the ground-state mixing of configurations based on the prolate and oblate
minima of the potential (cf. Fig.~\ref{fig:PES}). With the increase of angular momentum in the
ground-state band, e.g. $2^+_1$, $4^+_1$, etc., the states are progressively concentrated on the
prolate axis. For instance, $(\langle\beta\rangle,\langle\gamma\rangle) = (0.35,23^\circ)$ for $2^+_1$.
The average $\beta$-deformation in the ground-state band gradually increases because of centrifugal
stretching. Again we note that the empirical value $B(E2; 2^+_1\rightarrow 0^+_2)$ is accurately reproduced
by the present calculation using just the bare proton charge. In contrast to the SM prediction \cite{Force10},
the state $0^+_2$ is predominantly prolate, although one notices a relatively large
overlap between the wave functions of the states $0^+_1$ and $0^+_2$.
The mixing between these states is probably one of the reasons
for the high excitation energy of the second $0^+$ state, as predicted by the present calculation
(cf. Fig.~\ref{fig:spec-S44}). The probability distribution of the state $2^+_3$ is concentrated on the prolate
axis, and this state is connected by a strong transition to $0^+_2$: $B(E2; 2^+_3 \to 0^+_2) =
66$ $e^2{\rm fm}^4$, comparable to $B(E2; 2^+_1 \to 0^+_1)$.
We note, however, that for the ''coexisting" band based on
$0^+_2$ the calculated ratio $E(4^+)/E(2^+)$ is only 2.33.

The calculated second $2^+$ state displays a probability distribution extended in the
$\gamma$-direction and peaked on the oblate axis. As shown in
Fig.~\ref{fig:spec-S44} and Table~\ref{speS44}, this state is very close
to the experimental candidate for the $2^+_2$ state, which was suggested to be at
2335(39) keV by placing the 988 keV transition~\cite{Sohler02} on top of the $0^+_2$
or $2^+_1$ state~\cite{Force10}. The theoretical $2^+_2$ state can be
interpreted as the (quasi)-$\gamma$ band-head according to the strong E2
transitions to the states $3^+_1$ and $4^+_2$. For the three lowest $2^+$ states,
in Table~\ref{Kpro} we include the percentage of the $K=0$ and $K=2$ components
in the corresponding collective wave functions Eq.~(\ref{wave-coll}) (K denotes the
projection of the angular momentum on the intrinsic 3-axis), as well as the
spectroscopic quadrupole moments. The wave functions of the states $2^+_1$ and $2^+_3$
are dominated by $K=0$ components, and the spectroscopic quadrupole moments are
negative (prolate configurations) with comparable magnitudes. The positive quadrupole
moment of $2^+_2$ points to a predominant oblate configuration, and the $\approx 80\%$
contribution of the $K=2$ component in the wave function confirms that this state is the
band-head of a (quasi) $\gamma$-band (note the formation of the doublet
$3^+_1$ and $4^+_2$).

\begin {table}[htb]
\begin {center}
\caption{Percentage of the $K=0$ and $K=2$ components
(projection of the angular momentum on the body-fixed symmetry axis)
for the collective wave functions of the three lowest $2^+$ states in $^{44}$S,
and the corresponding spectroscopic quadrupole moments  (in $e ~{\rm fm}^2$).}
\bigskip
\begin {tabular}{c|ccc}
\hline
\hline
 & $K=0$ & $K=2$ & $Q_{\rm spec.}$ \\
\hline
$2^+_1$  & 88.4 & 11.6 & -10.9   \\
$2^+_2$  & 21.5 & 78.5 &   7.8   \\
$2^+_3$  & 80.0 & 20.0 &  -9.6   \\
\hline
\end{tabular}
\label{Kpro}
\end{center}
\end{table}
\begin{figure}[htb]
\includegraphics[scale=0.75]{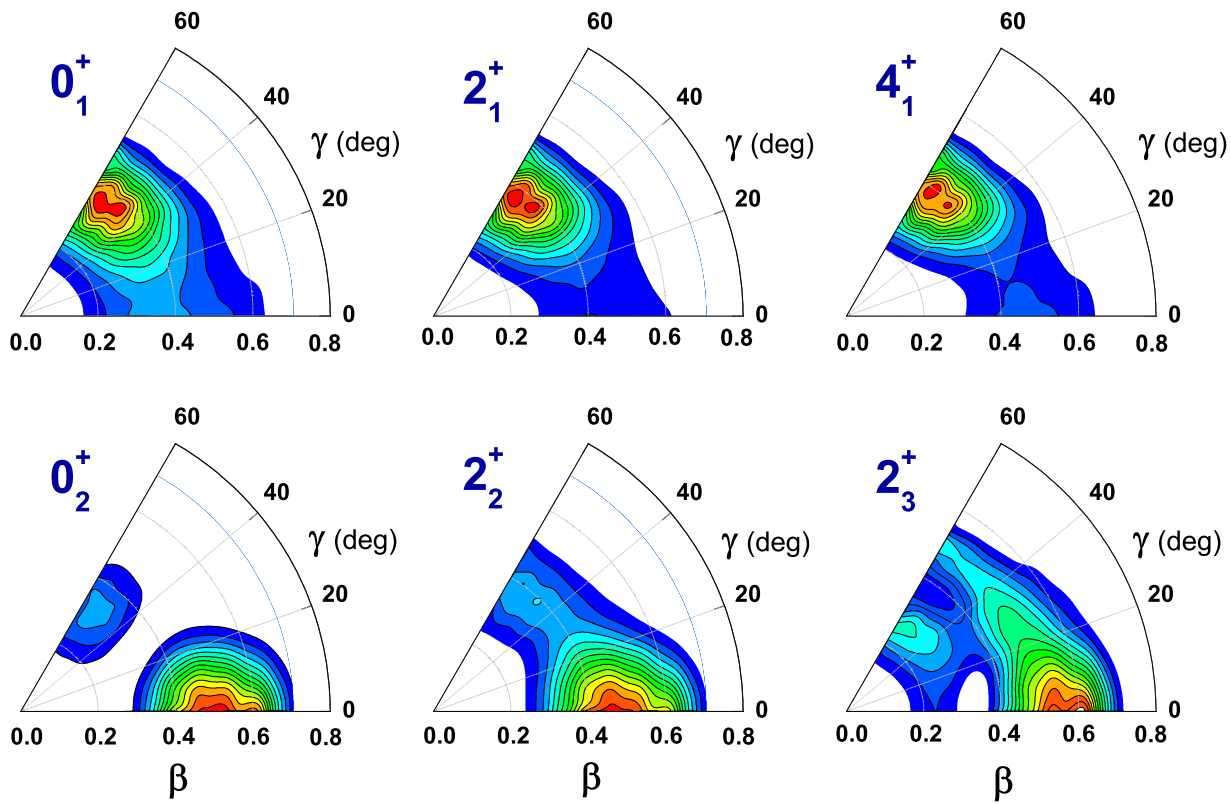}
\caption{\label{wavef-Si42}  (Color online)
Same as described in the caption to Fig. \ref{wavef-S44} but for the nucleus $^{42}$Si. }
\end{figure}

Finally, Fig.~\ref{fig:spec-Si42} shows the low-energy collective spectrum of $^{42}$Si.
Even though the excitation spectrum and transition pattern appear to be similar to that
of $^{44}$S, with the exception of a considerably weaker E2 transition
$0^+_2 \to 2^+_1$ (cf. Fig.~\ref{fig:spec-S44}), the ground-state band of this nucleus is
in fact based on the oblate minimum shown in the binding energy map of Fig.~\ref{fig:PES}.
This is nicely illustrated in Fig.~\ref{wavef-Si42} where, just like in the case of $^{44}$S
in Fig.~\ref{wavef-S44}, we plot the probability distributions of the collective wave functions
$0^+_1$, $2^+_1$, and $4^+_1$, the state $0^+_2$, and the two states
$2^+_2$ and $2^+_3$. The wave functions of the yrast states $0^+_1$, $2^+_1$,
and $4^+_1$ are concentrated along the
oblate axis. The state $0^+_2$ is strongly prolate deformed, with a peak in the
probability distribution at $\beta \approx 0.5$. This state has a much smaller overlap with
$2^+_1$ than in the case of $^{44}$S, and this explains the correspondingly weaker
transition. For $^{42}$Si, therefore, the solution of the collective Hamiltonian based on
the DD-PC1 functional, predicts a coexistence of the oblate yrast band and the prolate
sequence built on the strongly deformed state $0^+_2$. As already shown in Fig.~\ref{fig:obs},
the present calculation does not reproduce the exceptionally low excitation energy of
the state $2^+_1$: 770(19) keV \cite{Bastin07}. It is interesting, however, that the
calculated excitation energy of this state is very close to the SM prediction obtained using
the SDPF-NR effective interaction \cite{Bastin07}. Only by removing from the SDPF-NR
a schematic pairing Hamiltonian in the $pf$ shell, that is, by using the new effective
interaction SDPF-U \cite{Nowacki09}, the $2^+$ excitation energies of the silicon isotopes
can be brought in agreement with experiment.

\section{\label{secIV} Summary}

Structure phenomena related to the evolution of single-nucleon levels and shells in neutron-rich
nuclei present a very active area of experimental and theoretical research.
Among the microscopic models that can be used for a theoretical analysis of these
phenomena, the framework of nuclear energy density functionals (EDFs) presently provides a
complete and accurate description of ground-state properties and collective excitations
across the entire chart of nuclides. In this work we have used the recently introduced relativistic
EDF DD-PC1 \cite{NVR.08} to study the erosion of the $N=28$ spherical shell in neutron-rich nuclei
and the related phenomenon of shape evolution and shape coexistence in the $N=28$ isotones
$^{46}$Ar, $^{44}$S, $^{42}$Si, and $^{40}$Mg. Pairing correlations have been taken into account
by employing an interaction that is separable in momentum space, and is completely determined by
two parameters adjusted to reproduce the empirical bell-shaped pairing gap in symmetric
nuclear matter \cite{Niksic10}.

The N=28 shell closure is the first neutron magic number produced by the spin-orbit part of the
single-nucleon potential and, therefore, a relativistic mean-field model automatically
reproduces the N=28 spherical gap because it naturally includes the spin-orbit interaction
and the correct isospin dependence of this term, as it was shown more than ten years ago
in the axial RHB calculation of neutron-rich N=28 nuclei \cite{Lalazissis99}.
In particular, in the RMF approach there is no need for a tensor interaction to reproduce
the isospin dependence (quenching) of the spherical N=28 gap in neutron-rich nuclei,
as also shown in the present work in Table~\ref{shell}, compared to available data.

The functional DD-PC1 was adjusted exclusively to the experimental masses of a set of
64 deformed nuclei in the mass regions $A \approx 150-180$ and $A \approx 230-250$.
The present study of the $N=28$ isotones thus presents an extrapolation of DD-PC1 to
a completely different region of the nuclide chart, and a further test of the universality of
nuclear EDFs. It is not at all obvious that such an extrapolation
will produce results in agreement with experiment, especially in a detailed comparison
with spectroscopic data. The fact that it does is remarkable, and justifies the
approach to nuclear structure based on universal energy density functionals.

Starting from self-consistent binding
energy maps in the $\beta - \gamma$ plane, calculated in the relativistic Hartree-Bogoliubov
(RHB) model based on the functional DD-PC1, a recent implementation of the collective
Hamiltonian for quadrupole vibrations and rotations has been used to calculate the
excitation spectra and transition rates of $^{46}$Ar, $^{44}$S, $^{42}$Si, and $^{40}$Mg.
The parameters that determine the collective Hamiltonian: the vibrational inertial functions,
the moments of inertia, and the zero-point energy corrections, are calculated using the
single-quasiparticle energy and wave functions that correspond to each point on the
self-consistent RHB binding energy surface of a given nucleus. The diagonalization of the
collective Hamiltonian yields the excitation energies and wave functions used to calculate
various observables.

The calculation performed in this work has shown that the relativistic functional DD-PC1
provides an accurate microscopic interpretation of the strong reduction of the $N=28$
spherical energy gap in neutron-rich nuclei, and a quantitative description of the evolution
of shapes in $N=28$ isotones in terms of single-nucleon orbitals as functions of the
quadrupole deformation parameters $\beta$ and $\gamma$. In particular, the predicted
values for the spherical shell gap in $^{48}$Ca (4.73 MeV) and in $^{46}$Ar (4.48 MeV),
are very close to the data: 4.80 MeV in $^{49}$Ca and 4.47 MeV in $^{47}$Ar. The solutions
of the collective Hamiltonian based on DD-PC1 reproduce the evolution with proton
number of characteristic collective observables the excitation energy of the first $2^+$
state, the ratio $E(4^+_1)/E(2^+_1)$, and the reduced transition probability
B(E2; $2^+_1 \to 0^+_1$). In the present work we have focused on $^{44}$S,
for which recent data point towards a coexistence of shapes with different deformations
in the low-energy excitation spectrum. It has been shown that the formation
of the oblate neutron and prolate proton gaps, illustrated in Fig.~\ref{44S}, is at
the origin of the predicted coexistence of deformed shapes in $^{44}$S.
The spectroscopic results have been compared to available data,
to triaxial (collective Hamiltonian) and axial (generator coordinate method) calculations
based on the Gogny D1S HFB self-consistent mean-field energy maps, and to recent
shell-model (SM) calculations using the new SDPF-U effective interaction. The present
results are in qualitative agreement with previous calculations based on the Gogny D1S HFB
model and, in particular, reproduce the data on both the excitation energy of the first excited
state $2^+_1$ and the reduced transition probability $B(E2; 2^+_1 \to 0^+_1)$, and the
theoretical value for $B(E2; 0^+_2 \to 2^+_1)$ is also in good agreement with data.
The experimental ratio $B(E2; 2^+_1 \to 0^+_1) / B(E2; 2^+_1 \to 0^+_2)$ is 7.5, and the
calculated value is 5.2. The theoretical monopole transition strength
$\rho^2(E0; 0^+_2\rightarrow 0^+_1) (\times 10^3)$ = 23 is somewhat larger than the
corresponding experimental value of 8.7(7). The calculation of transition rates in the
collective Hamiltonian model is completely parameter-free.
One might notice that the results predicted by the functional DD-PC1 have been
compared to those obtained using effective interactions that were fine-tuned to data
that include also this mass region, or adjusted exclusively to data in this region of the
mass table (shell-model interactions). The fact that a global density functional
can even compete in a spectroscopic calculation with shell-model interactions
specifically customized to this mass region, and the level of agreement with experiment,
presents a valuable result.

A discrepancy with respect to experiment in $^{44}$S is the high excitation energy
predicted for the state $0^+_2$, a factor of two compared to data. It appears that the model
predicts too much mixing between the two lowest $0^+$ states, and this leads to an
enhancement of the corresponding monopole transition strength. The pronounced mixing
between the calculated $0^+_1$ and $0^+_2$ states, and the resulting repulsion, could be at
the origin of the high excitation energy of $0^+_2$. The most obvious reason
is that this is an intrinsic prediction of the functional DD-PC1. To check this one would have
to perform calculations using different functionals \cite{Zhao10}. However, since also
the Gogny D1S + 5DCH model yields a similar result, the functional itself probably  is not the main
problem. A more probable reason is that the mass parameters calculated in the cranking approximation
are simply too small, as discussed in Sec.~\ref{ssecIII}. Finally, the excited $0^+$ state could also have
pronounced non-collective components that are not included in our model space (2-quasiparticle contributions).
This is certainly a possibility, and it would partially explain why the calculated B(E2) to the first $2^+$ state is larger
than the experimental value. The shell-model calculation of Ref.~\cite{Force10} predicts the excitation energy of
$0^+_2$ in better agreement with experiment, but the calculated B(E2) for the transition from the
first $2^+$ state is more than a factor two larger than the experimental value
(only about 50\% larger in the present calculation). Therefore, it appears that the structure of the
second $0^+$ state in $^{44}$S remains an open problem.

This present analysis of low-energy spectra of $N=28$ isotones has clearly demonstrated the
advantages of using EDFs in the description of deformed nuclei: an intuitive mean-field interpretation
in terms of coexisting intrinsic shapes and the evolution of single-particle states, spectroscopic calculations
performed in the full model space of occupied states, and the universality of EDFs that enables
their applications to nuclei in different mass regions, including short-lived systems far from stability.

\begin{acknowledgements}
This work was supported in part by the Major State 973 Program 2007CB815000,
the NSFC under Grant Nos. 10975008, 10947013, 11105110, and 11105111,
the Southwest University Initial Research
Foundation Grant to Doctor (Nos. SWU110039, SWU109011), the Fundamental
Research Funds for the Central Universities (XDJK2010B007 and XDJK2011B002)
and the the MZOS - project 1191005-1010.
The work of J.M, T.N., and D.V. was supported in part by
the Chinese-Croatian project "Nuclear structure and astrophysical applications".
T. N. and Z. P. Li acknowledge support by the Croatian National Foundation for
Science.
\end{acknowledgements}
\clearpage


\end{document}